# Tunneling spectra of $TaO_x$ junctions for van der Waals superconductors

Yixuan Niu[1,†], Jun Cheng[1,†], Shiji Ding[1,†], Zhongxin Guo[1], Shang Wang[1], Chenglong Li[1], Meining Zhang[2] and Peng Cai[1,*]

*[1] School of Physics and Key Laboratory of Quantum State Construction and Manipulation (Ministry of Education), Renmin University of China, Beijing 100872, China*

*[2]School of Chemistry and Life Resources, Renmin University of China, Beijing 100872, China*

[†] These authors contributed equally to this work.

[*] Correspondence should be addressed to P.C. (pcai@ruc.edu.cn).

**Tunneling spectroscopy and its evolution are crucial for elucidating the intricate electronic structure and emergent phenomena in quantum materials. Nevertheless, high-quality measurements—specifically those tracking evolution across temperature and external fields—remain a formidable challenge. We have fabricated a high-quality $TaO_x$-based planar tunneling junction by using magnetron sputtering for van der Waals (vdW) superconductors. Using the vdW superconductor $Bi_2Sr_2CaCu_2O_{8+\delta}$ (Bi2212) as a benchmark, this platform yields high-quality tunneling spectra, reproducing the electronic signatures obtained from scanning tunneling spectra acquired from atomically clean surfaces under ultra-high vacuum conditions. This architecture enables high-precision spectroscopy across extensive temperature and magnetic field ranges, offering a universal strategy for probing the electronic structures of diverse two-dimensional systems and facilitating future explorations of material properties.**

The determination of the intertwined electronic ground state [1-4], alongside the evolution of the electronic states across diverse phase transitions, remains central to decoding the enigmatic mechanism of unconventional superconductivity in quasi two-dimensional (2D) materials. By facilitating the quantum tunneling of electrons across a barrier, tunneling spectroscopy provides a direct probe of the energy gap [5], quasiparticle excitations, and many-body interactions. The advent of scanning tunneling spectroscopy (STS) [6] has revolutionized the ability to probe quasi 2D materials, enabling the realization of pristine, controllable tunneling junctions under ultra-high vacuum with energy and spatial precision. As illustrated in Fig. 1(a), upon applying a bias voltage $V$ to the sample, electrons tunnel into the empty state of the sample from the metal with a nearly constant density of states (DOS) around Fermi level, such that the differential conductance is proportionally with the sample's DOS. Figure 1(b) shows STS operation via vacuum barrier tunneling between a tip apex and individual atom on the sample surface. While powerful, the high cost of STS instrumentation, combined with the extreme sensitivity of the tip–sample junction to thermal gradient and mechanical vibrations [7], severely restricts the experimental throughput and versatility of this technique. Consequently, for probing emergent phenomena under diverse physical conditions (variable temperature and strong magnetic field), acquiring high-fidelity differential conductance spectra demands exceptional tunneling junction stability, which remains a formidable experimental challenge.

As a robust alternative to STS techniques, planar junctions provide a device-based platform capable of maintaining stable tunneling junction under diverse conditions [8]. Planar tunneling junctions typically feature a sandwich heterostructure as shown in Fig. 1(c), wherein electrons tunnel vertically through an insulating barrier across a macroscopic interface, facilitating collective electron tunneling from the metallic electrode into the target sample. Notably, planar tunneling spectroscopy offers distinct advantages over STM: the robust insulating barrier renders the device architecture resilient to mechanical vibrations and environmental noise, thereby this technique significantly facilitates continuous, high-precision measurements across expansive temperature and magnetic field regimes [8]. Furthermore, unlike point-contact spectroscopy, which often suffers from interfacial instability during thermal cycling, the planar tunneling architecture offers a non-destructive, robust alternative. The well-defined barrier layer ensures reliable tunneling contact, thereby minimizing the measurement uncertainties typically associated with mechanical point-contact variations [9-10].

Historically, however, the proliferation of planar junctions was impeded by the formidable technical hurdles in fabricating high-quality tunneling barriers, where issues such as interfacial roughness, structural disorder, and pinhole defects frequently failed the junction and compromised spectroscopic fidelity [11-12]. Although conventional planar tunneling junctions can prepare high-quality insulating barriers with techniques such as pulsed laser deposition (PLD) [13] or molecular beam epitaxy (MBE) [14], the choice of barrier materials must strictly meet lattice matching and other conditions, which limits their application to vdW superconductor. Recently, ultra-high quality $AlO_x$ barrier grown by atomic layer deposition (ALD) is attempted for planar tunneling junction for vdW superconductor, however, the junction based $AlO_x$ require narrow thickness window and uniformity, rendering the fabrication of high performance $AlO_x$ junction exceptionally challenging for vdW superconductors. Tantalum oxide $TaO_x$ emerges as a superior tunneling barrier candidate, distinguished by its high chemical stability and density, making it an ideal barrier for high-performance tunneling junctions [15]. Unlike conventional insulators such as $AlO_x$ and MgO [16-17], which necessitate extreme thinness to achieve the desired tunneling barrier height, $TaO_x$ functions effectively at greater physical thicknesses [15]. Consequently, $TaO_x$-based junction is promising to exhibit an improved tolerance to thickness variations, ensuring superior device reproducibility across vdW heterostructures, however remain to be experimentally verified.

Here we present a robust fabrication protocol for $TaO_x$-based planar tunneling junctions for van der Waals (vdW) superconductors, integrating magnetron sputtering of barrier heterostructure with advanced transfer technology for two-dimensional materials. Furthermore, by performing vdW material exfoliation [18] and dry transfer within an inert atmosphere, we effectively minimize interfacial contamination, ensuring the formation of pristine, well-defined junctions. To validate the versatility and reliability of this platform, we applied it to the high transition temperature superconductor Bi2212, a prototypical vdW material. The tunneling spectra obtained from low-temperature measurements are highly consistent with the STS results of similar samples [2]. As the temperature increases, the gap signature of coherent SC state gradually weakens and disappears around $T_c$, while a pseudogap can be observed far above $T_c$ even close to room temperature. Upon applying a magnetic field, the superconducting gap is suppressed while the pseudogap remains almost unchanged.

Figure 1(d) shows the schematic protocol to achieve the $TaO_x$ based planar tunneling junction for vdW superconductor. Gold contact electrodes on insulating substrate were patterned using masks and deposited via thermal evaporation. Ta electrode was aligned in between the gold electrodes and deposited using magnetron sputtering. Subsequently, $TaO_x$ was formed as a stable insulating layer for tunneling barrier. After these electrodes were prepared, the thin flake of vdW superconductor is exfoliated and transferred onto $TaO_x$ electrode. The architecture of the device is schematically depicted in the side-view illustration in Fig. 1(e), revealing a planar tunneling junction composed of a flake-$TaO_x$-Ta heterostructure. Given that Bi2212 is highly susceptible to contamination and degradation in air, the entire process of exfoliation and transferring was carried out in an inert gas atmosphere. Meanwhile, dry transfer method [19] was employed to maximize the structural integrity and surface cleanliness of the samples. The optical image of Bi2212 device for the tunneling spectroscopy is shown in Fig. 1(f). Moreover, the integration of four gold electrodes enables simultaneous acquisition of tunneling spectra and four-probe resistance measurements. Differential conductance measurements were performed using standard lock-in amplification techniques with frequency = 137Hz.

To evaluate the performance of the tunneling junction, the Bi2212 device was cooled down to a base temperature of 1.5 K after preparation. Figure 2(a) shows the tunneling characteristic of the junction at base temperature. The I-V curve exhibits a clear nonlinear behavior, consistent with the tunneling model described by the BTK theory [10]. As shown in Fig. 2(b), the bias dependence of dI/dV curve, which is proportional to DOS [20]. The V-shaped gap structure with symmetric coherent peak can be attributed to the d-wave pairing symmetry of cuprate superconductors. The finite DOS at zero bias voltage may be due to integration of the defect state and thermal excitation of d-wave quasiparticles. The superconducting coherence peaks of Bi2212 appear at ±47 meV, which is in good agreement with the typical superconducting energy gap $\Delta_{SC} \approx 40$ meV obtained from scanning tunneling microscopy studies of similar Bi2212 samples [21].In addition to the superconducting coherence peaks and the superconducting energy gap, the spectrum also shows two typical features: one is the left-high and right-low tilted asymmetry of the background outside the energy gap; the other is a distinct dip structure on the negative bias side outside the superconducting coherence peaks.

To further explore the physical origin of this spectral shape, we systematically measured the evolution of the differential conductance spectrum with temperature. As shown in Fig. 2(c), the dI/dV curve was tracked over a temperature range from 1.5 K to 220 K. As the temperature increased, the intensity of the superconducting coherence peaks gradually decreased, the DOS at zero-bias gradually increased, and the dip feature outside the peaks are weakened. Overall, the superconducting energy gap features became shallower at higher temperatures. When the temperature exceeded the superconducting transition temperature $T_c$ = 94 K, the dI/dV curve tended to flatten and no abrupt changes were observed. Above $T_c$, the coherence peaks disappeared, but the DOS suppression still existed, corresponding to the typical pseudogap region. The dip structure disappeared along with the coherence peaks, and a broad hump appeared on the positive bias side, consistent with previous STS results. The pseudogap was still observable up to 220 K, although its characteristics gradually weakened. Only at about 250 K did the energy gap and hump structure become indistinguishable from the background. Similar temperature-dependent STS experiments have been conducted in cuprate superconductors [22], but the data is complicated for extraction of precise electronic signatures since substantial noise was introduced by thermal vibration and drift at high temperatures.

To get further insight on the evolution of electronic state, Fig. 2(d) shows the temperature dependence of dI/dV at zero bias voltage. Due to the thermal expansion of the tunneling junction, the measured dI/dV slightly decreases with increasing temperature above 200 K. Such measurements have not yet been realized in STS measurement. In the resistance-temperature (R-T) curves, an inflection point appears near $T_c$ during the process of dropping to zero resistance, and decrease rate significantly accelerates at lower temperature ranges. Figure 2(e) shows the derivatives of the two curves in Fig. 2(d), it can be seen that the main peak of the temperature dependence of dI/dV curve appears at about 80 K, which is slightly lower than the superconducting transition temperature $T_c$ = 94 K determined from the R-T curve, while the secondary peak corresponds to the superconducting transition temperature. It is generally believed that the dI/dV curve of cuprate superconductors evolves continuously across $T_c$ without obvious abrupt changes. However, our experimental results show that the electronic DOS at the Fermi level exhibits significant changes near $T_c$, providing experimental evidence for understanding the physical nature of the superconducting phase transition in copper-based superconductors. This feature is consistent with results reported in the existing literature [8].

Figure 2(f) shows the tunneling spectra obtained at several typical temperatures within the range of 1.5K to 88K below the superconducting transition temperature. They share almost similar background at high bias voltage, revealing the excellent quality of the tunneling junction. To effectively isolate the superconducting features and eliminate the normal-state background, each tunneling spectrum was normalized to the reference spectrum acquired at 94 K, as shown in Fig. 2(g). The coherence peak positions exhibit the particle-hole symmetry characteristic of superconducting state, with the normalized spectra clearly resolving a gap structure consistent with *d*-wave pairing symmetry [2,23].

Figure 3 demonstrates another advantage of the planar tunneling spectra, the robustness in capturing spectra across an extensive range of magnetic fields. At base temperature, the superconducting coherence peaks undergo suppression upon the application of a magnetic field, accompanied by a filling of the zero-bias conductance. It can be naturally attributed to the emergence of vortex in Bi2212. Previous scanning tunneling microscopy studies on Bi2212 have shown [24] that the coherence peaks at the vortex core centers are completely suppressed, and low-energy bound states appear within the energy gap. The partial suppression of the superconducting coherence peaks at 8 T is attributed to the limited volume fraction occupied by vortex cores; the surrounding regions remain in the superconducting state and dominatingly contribute to the tunneling spectra. Meanwhile, the dip at negative bias is significantly weakened in a magnetic field, indicating that this feature is closely related to the superconducting state.

At elevated temperatures below $T_c$, the suppression of the magnetic field on superconductivity follows the similar trend as at base temperature, but with significantly reduced intensity, as shown in Fig. 3(b,c). When the temperature rises to 100 K (slightly above $T_c$), as shown in Figure 3(d), the magnetic field effect on the dI/dV spectrum almost completely disappears, which is another characteristic change marking the transition from the superconducting phase to the pseudogap phase. These observations are consistent with low-temperature STS experiments on $Bi_2Sr_2CuO_{6+\delta}$ (Bi2201), indicating that upon a magnetic field, the pseudogap remains almost unchanged while the superconducting gap is continuously suppressed [25].

To evaluate the robustness and reproducibility of the junction architecture for tunneling spectroscopy, a systematic comparative analysis is performed on different junctions based on Bi2212. As revealed in Fig. 4(a), the tunneling spectra with $TaO_x$ junctions is compared with previously reported data for Bi2212 devices utilizing $AlO_x$ and vacuum barriers. This consistence highlights the effectiveness of the architecture based on $TaO_x$ architecture for vdW superconductors. The minor variations in the gap values are primarily attributed to two intrinsic factors of Bi2212. First, subtle variations in oxygen stoichiometry within the bulk crystals result in slight variation of SC gap size. Second, the intrinsic surface sensitivity of Bi2212, particularly concerning oxygen-loss at room temperature, renders the tunneling spectra measurement susceptible to the interfacial preparation history. To validate later factor, we carried experiments on another freshly prepared junction under reduced temporal exposure for comparison as revealed in Fig. 4(b). Cryogenic measurements performed immediately post-transferring Bi2212 (within half day) yield a smaller gap compared to sample subjected to prolonged aging. This observation aligns with the well-established doping dependence of Bi2212, in which a decrease in oxygen stoichiometry suppresses the hole carrier concentration, subsequently leading to an increase in the SC gap size. The tunneling junction architecture is robust for different Bi2212 flakes with slightly different doping. The robustness of the tunneling junctions can be further demonstrated by the persistence of reproducible tunneling spectra as shown in Fig. 4(c) even after prolonged storage at room temperature, confirming the long-term stability of the $TaO_x$-based architecture for vdW superconductors. The evolution towards large gap can be again attributed to the gradual depletion of oxygen from the Bi2212 surface at room temperature, which drives the surface stoichiometry towards an underdoped state.

In summary, we have established a robust platform for planar tunneling junctions for vdW superconductors by integrating magnetron sputtering for tunneling barrier and transfer method for thin flakes. The reliability of this architecture is validated by spectral consistency between our tunneling measurements on exfoliated Bi2212 and established STS data. By enabling spectroscopic studies across extensive temperature and magnetic field ranges, this approach provides a powerful experimental avenue for probing the SC gap evolution and phase transition in vdW superconductors. Furthermore, the modular nature of this technique permits the characterization of a wide spectrum of vdW systems, ranging from conventional superconductors to complex twisted stacks and cuprates with various doping. This planar tunneling architecture is inherently compatible with extreme environments, enabling versatile spectroscopic characterization of vdW superconductor from ambient conditions to high-pressure regimes via diamond anvil cell integration [26], thereby providing a promising avenue for probing their pressure-tuned electronic structures and mapping comprehensive electronic phase diagrams.

**Acknowledgements:** This work was supported by National Key Research Program of China (2022YFA1403102), the Innovation Program for Quantum Science and Technology (grant No. 2021ZD0302502), the National Natural Science Foundation of China (Grant No. 12074424). Peng Cai is supported by the Fundamental Research Funds for the Central Universities, and the Research Funds of Renmin University of China.

## Figure Captions:

**Figure 1 Schematic of the planar tunneling junction fabrication and tunneling spectroscopy.** **(a)** Schematic of DOS measurement via tunneling spectroscopy. **(b)** Schematic of the vacuum-based tunneling junction configuration for STS measurements. **(c)** Electron tunneling from a metal electrode to the sample in an insulator-based planar junction. **(d)** Fabrication workflow of planar tunneling junction devices: sequential deposition of gold contact electrodes and a Ta tunneling electrode, followed by controlled formation of the $TaO_x$ tunneling barrier and precise transfer of exfoliated flakes onto the electrodes. Interface cleanliness and device structural integrity is attempted to preserve in this protocol. **(e)** Cross-sectional view of the device architecture, illustrating the electron tunneling path from the metal electrode through the $TaO_x$ barrier into the 2D vdW flake. **(f)** Optical image of a representative Bi-2212 device (device #1) and a schematic of the tunneling measurement configuration.

**Figure 2 Temperature dependence of the tunneling spectra of near-optimally doped Bi-2212 device#1.** **(a)** Current-voltage characteristic of the device at 1.5 K. **(b)** Tunneling spectrum corresponding to **(a)**. **(c)** Temperature dependence of tunneling spectra. For clarity, each tunneling spectrum is shifted vertically by 0.1 μS except for the 1.5 K curve. **(d)** Temperature dependence of the zero-bias conductance (red) and the device resistance (blue). **(e)** Derivatives of the two curves corresponding to **(d)**. **(f)** Tunneling spectra at representative temperatures below $T_c$, showing a consistent background at high energies. **(g)** Normalized spectra obtained by dividing the tunneling data in **(f)** by the normal-state spectrum at 94 K, unveiling the intrinsic superconducting features.

**Figure 3 Effect of an out-of-plane magnetic field on the tunneling spectra of a nearly optimally doped Bi-2212.** **(a–d)** Tunneling spectra of Bi-2212 (device #1) at **(a)** 1.5 K, **(b)** 40 K, **(c)** 85 K, and **(d)** 100 K, measured under zero field (blue) and an 8 T magnetic field (red).

**Figure 4 Validation, reproducibility, and stability of nearly optimally doped Bi-2212 devices.** **(a)** Comparison of the tunneling spectra obtained in this work with previous results from STS and $AlO_x$-based planar tunneling junctions. **(b)** Reproducibility test shows consistent tunneling characteristics across different devices. **(c)** Stability test for device #1. The second measurement was performed after exposing the device to ambient air at room temperature for three days,

showing slight degradation. All spectra have been normalized to their respective normal-state tunneling conductance.

## Reference:

[1] B. Keimer, S. A. Kivelson, M. R. Norman, et al. From quantum matter to high-temperature superconductivity in copper oxides. *Nature* **518**, 179–186 (2015).

[2] Ø. Fischer, M. Kugler, I. Maggio-Aprile, et al. Scanning tunneling spectroscopy of high-temperature superconductors. *Rev. Mod. Phys.* **79**, 353–419 (2007).

[3] P. Cai, X. Zhou, W. Ruan, et al. Visualizing the microscopic coexistence of spin density wave and superconductivity in underdoped $NaFe_{1-x}Co_xAs$. *Nat. Commun*. **4**, 1596 (2013).

[4] X. D. Zhou, P. Cai, A. F. Wang, W. Ruan, C. Ye, X. H. Chen, Y. Z. You, Z. Y. Weng, Y. Y. Wang. Evolution from unconventional spin density wave to superconductivity and a novel gap-like phase in $NaFe_{1-x}Co_xAs$. *Phys. Rev. Lett*. **109**, 037002 (2012).

[5] I. Giaever. Energy Gap in Superconductors Measured by Electron Tunneling. *Phys. Rev. Lett*. **5**, 147–148 (1960).

[6] H. Maeda, Y. Tanaka, M. Fukutomi, T. Asano. A New High-$T_c$ Oxide Superconductor without a Rare Earth Element. *Jpn. J. Appl. Phys.* **27**, L209–L212 (1988).

[7] M. Okano, K. Kajimura, S. Wakiyama, F. Sakai. Vibration isolation for scanning tunneling microscopy. *J. Vac. Sci. Technol.* **A 5**, 3366–3370 (1987).

[8] Y. Ji, H. Wang, Z. Dong, et al. Planar tunneling spectroscopy on van der Waals superconductors with $AlO_x$ junction grown by atomic layer deposition. *J. Appl. Phys*. **133**, 013903 (2023).

[9] N. Miyakawa, P. Guptasarma, J. F. Zasadzinski, et al. Strong dependence of the superconducting gap on oxygen doping from tunneling measurements on $Bi_2Sr_2CaCu_2O_{8+\delta}$. *Phys. Rev. Lett.* **80**, 1 (1998).

[10] Ge He, Zhong-Xu Wei, et al. Distinction between critical current effects and intrinsic anomalies in the point-contact Andreev reflection spectra of unconventional superconductors. *Chin. Phys. B* **27**, 047403 (2018).

[11] R. Jansen and J. S. Moodera. Influence of barrier impurities on the magnetoresistance in ferromagnetic tunnel junctions. *J. Appl. Phys.* **83**, 6682 (1998).

[12] D. Allen, R. Schad, G. Zangari, I. Zana, M. Tondra, D. Wang, and D. Reed. Comparison of defect density measurements in magnetic tunnel junctions. *J. Appl. Phys*. **89**, 6662 (2001).

[13] C. Richter, H. Boschker, W. Dietsche, et al. Interface superconductor with gap behaviour like a high-temperature superconductor. *Nature* **502**, 528–531 (2013).

[14] P. Zhou, L. Chen, I. Sochnikov, et al. Tunneling spectroscopy of c-axis epitaxial cuprate junctions. *Phys. Rev. B* **101**, 224512 (2020).

[15] P. Rottländer, M. Hehn, O. Lenoble, et al. Tantalum oxide as an alternative low height tunnel barrier in magnetic junctions. *Appl. Phys. Lett.* **78**, 3274 (2001).

[16] S. Yuasa, T. Nagahama, A. Fukushima, et al. Giant room-temperature magnetoresistance in single-crystal Fe/MgO/Fe magnetic tunnel junctions. *Nat. Mater*. **3**, 868–871 (2004).

[17] I. Giaever. Electron Tunneling Between Two Superconductors. *Phys. Rev. Lett*. **5**, 464–466 (1960).

[18] K. S. Novoselov, A. K. Geim, S. V. Morozov, et al. Two-dimensional atomic crystals. *Proc. Natl. Acad. Sci. U.S.A.* **102**, 10451–10453 (2005).

[19] A. Castellanos-Gomez, M. Buscema, R. Molenaar, et al. Deterministic transfer of two dimensional materials by all-dry viscoelastic stamping. *2D Mater*. **1**, 011002 (2014).

[20] J. Bardeen. Tunnelling from a Many-Particle Point of View. *Phys. Rev. Lett*. **6**, 57–59 (1961).

[21] C. Renner, O. Fischer. Vacuum tunneling spectroscopy and asymmetric density of states of $Bi_2Sr_2CaCu_2O_{8+\delta}$. *Phys. Rev. B* **51**, 9208–9211 (1995).

[22] A. Matsuda, S. Satoshi, et al. Temperature and doping dependence of the $Bi_{2.1}Sr_{1.9}CaCu_2O_{8+\delta}$ pseudogap and superconducting gap. *Phys. Rev. B* **60**, 1377 (1999).

[23] Z. X. Shen, D. S. Dessau, B. O. Wells, et al. Anomalously large gap anisotropy in the a-b plane of $Bi_2Sr_2CaCu_2O_{8+\delta}$. *Phys. Rev. Lett.* **70**, 1553–1556 (1993).

[24] S. H. Pan, E. W. Hudson, A. K. Gupta, et al. STM studies of the electronic structure of vortex cores in $Bi_2Sr_2CaCu_2O_{8+\delta}$. *Phys. Rev. Lett.* **85**, 1536 (2000).

[25] Y. He, Y. Yin, M. Zech, A. Soumyanarayanan, M. M. Yee, T. Williams, et al. Fermi Surface and Pseudogap

Evolution in a Cuprate Superconductor. *Science* **344**, 608–611 (2014).
[26] L. G. Pimenta Martins, R. Comin, M. J. S. Matos, et al. High-pressure studies of atomically thin van der Waals materials. *Appl. Phys. Rev.* **10**, 011313 (2023).

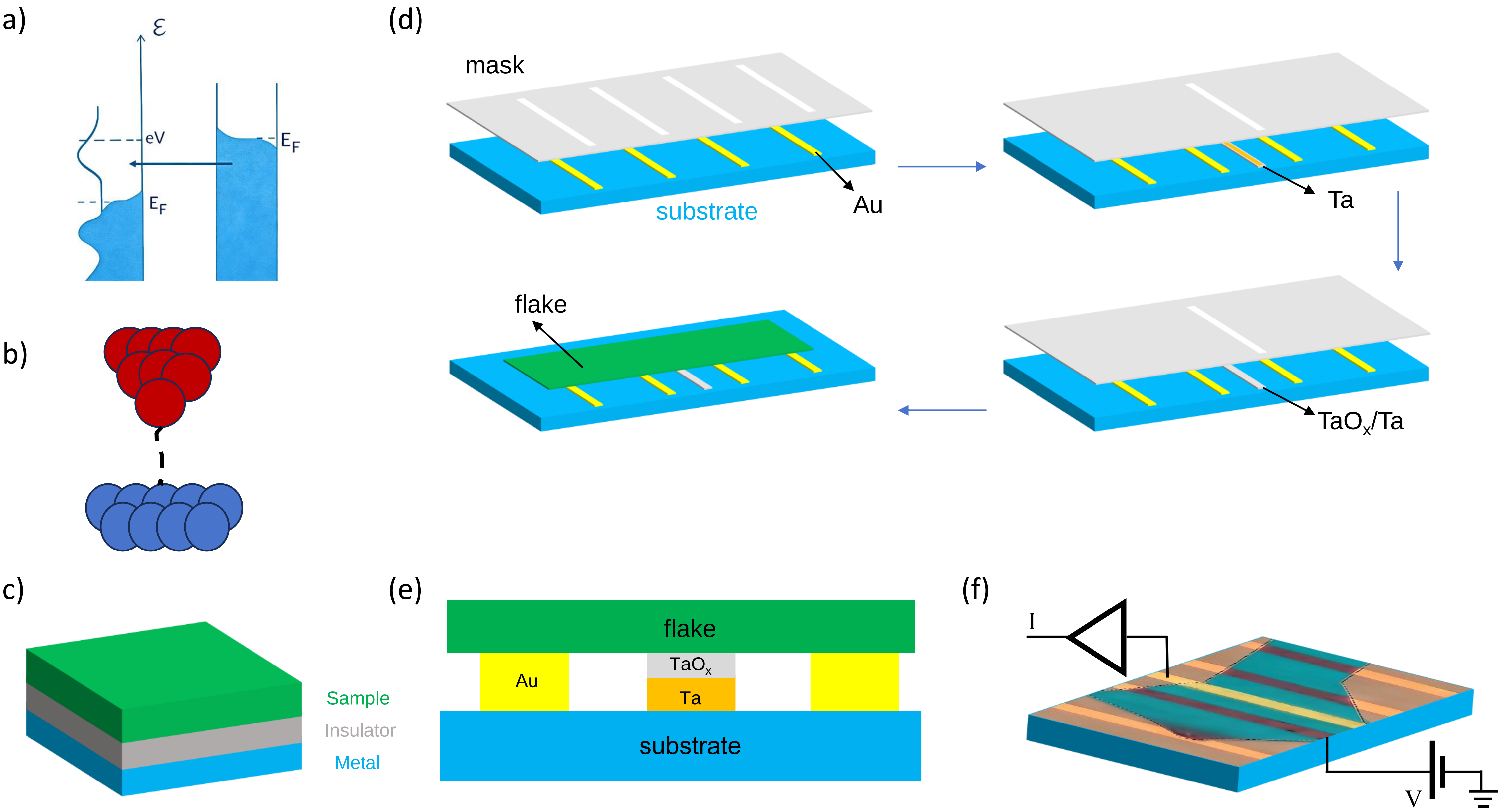
(a)
ℰ
eV
E_F
E_F
(b)
(c)
Sample
Insulator
Metal
(d)
mask
substrate
Au
Ta
flake
TaO_x/Ta
(e)
flake
Au
TaO_x
Ta
substrate
(f)
I
V

Fig.1

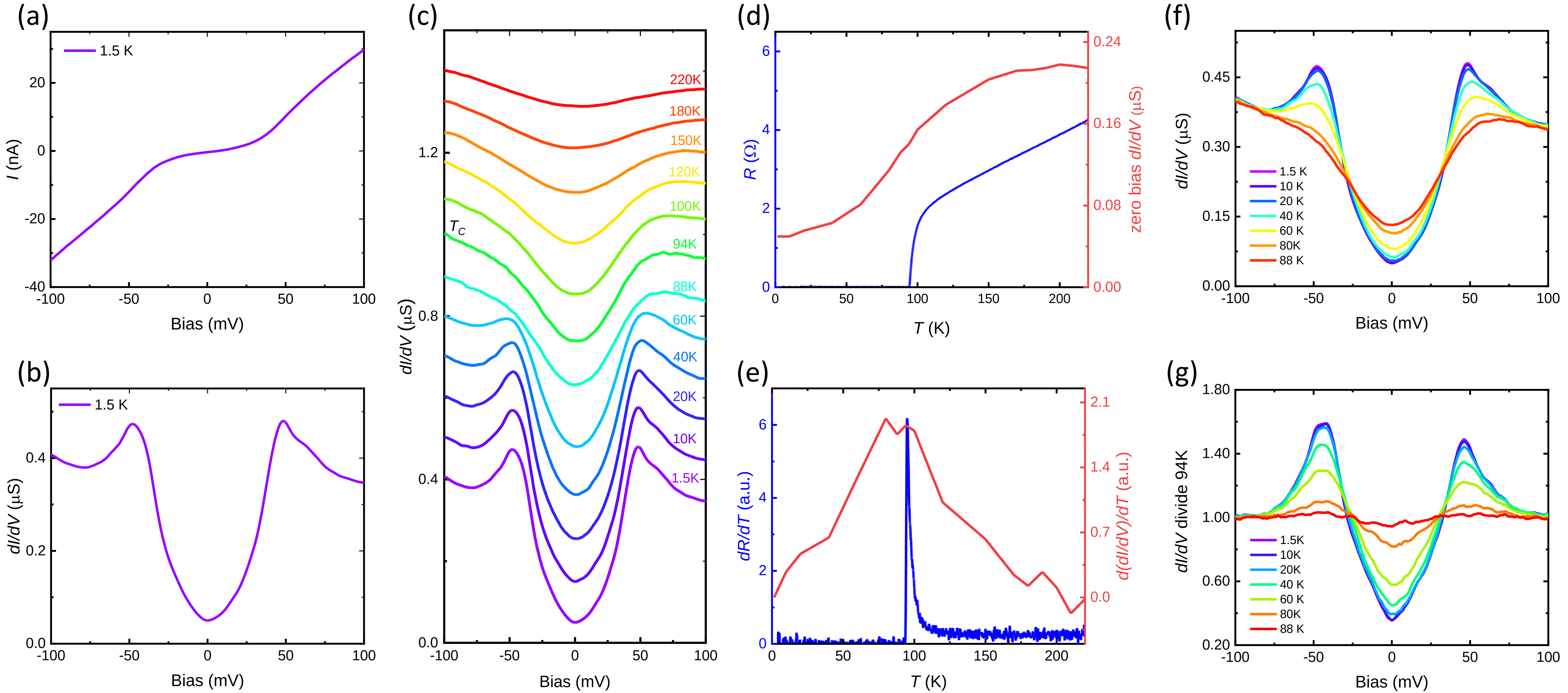

(a)
1.5 K
I (nA)
Bias (mV)
(b)
1.5 K
dI/dV (μS)
Bias (mV)
(c)
220K
180K
150K
120K
100K
94K
88K
60K
40K
20K
10K
1.5K
T_C
dI/dV (μS)
Bias (mV)
(d)
R (Ω)
zero bias dI/dV (μS)
T (K)
(e)
dR/dT (a.u.)
d(dI/dV)/dT (a.u.)
T (K)
(f)
1.5 K
10 K
20 K
40 K
60 K
80K
88 K
dI/dV (μS)
Bias (mV)
(g)
1.5K
10K
20K
40 K
60 K
80K
88 K
dI/dV divide 94K
Bias (mV)

Fig.2

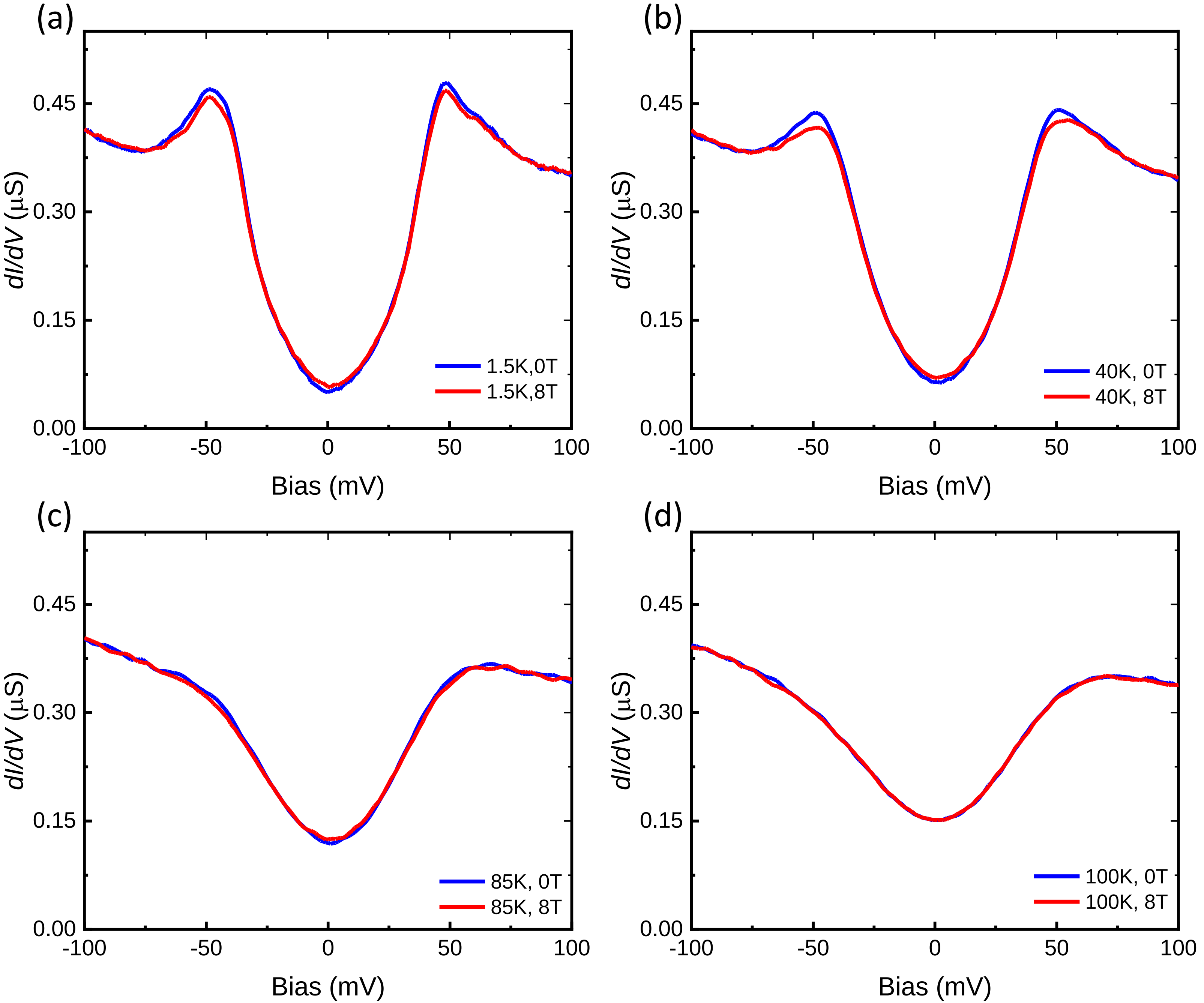
(a)
(b)
(c)
(d)
dI/dV (μS)
Bias (mV)
0.00
0.15
0.30
0.45
-100
-50
0
50
100
1.5K,0T
1.5K,8T
40K, 0T
40K, 8T
85K, 0T
85K, 8T
100K, 0T
100K, 8T

Fig.3

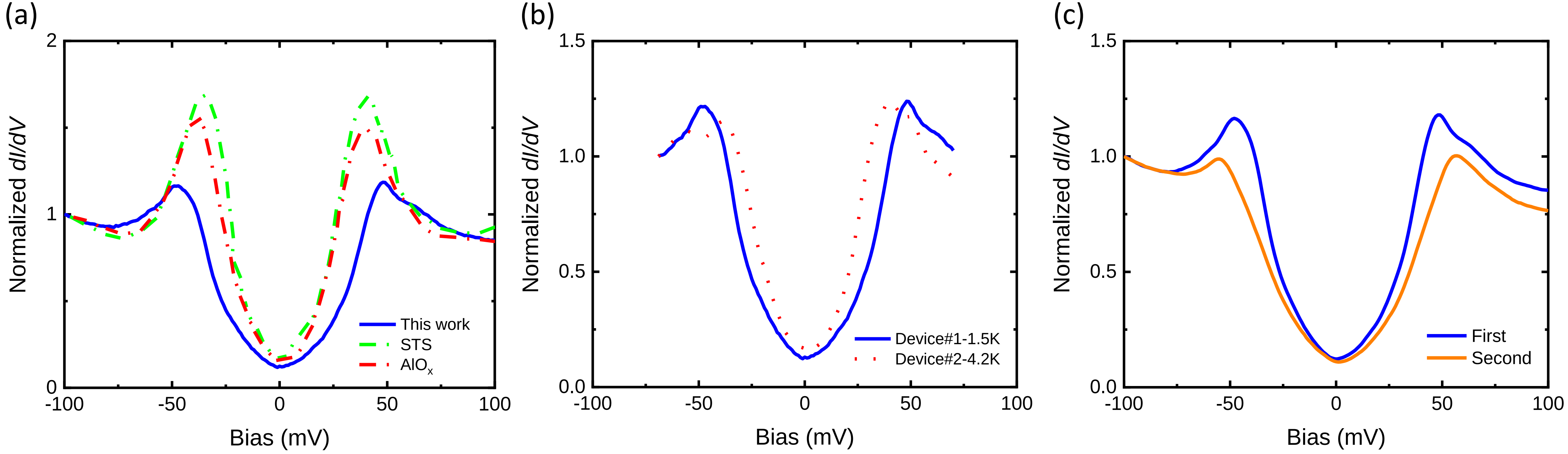
(a)
2
1
0
Normalized dI/dV
-100
-50
0
50
100
Bias (mV)
This work
STS
AlOx
(b)
1.5
1.0
0.5
0.0
Normalized dI/dV
-100
-50
0
50
100
Bias (mV)
Device#1-1.5K
Device#2-4.2K
(c)
1.5
1.0
0.5
0.0
Normalized dI/dV
-100
-50
0
50
100
Bias (mV)
First
Second

Fig.4